\documentclass{aastex63}

\usepackage{xspace}

\newcommand{\RomanSpelled}{{\it Nancy Grace Roman Space Telescope}\xspace}
\newcommand{\RST}{{\it Roman}\xspace}
\newcommand{\HST}{{\it HST}\xspace}
\newcommand{\ang}{\AA\xspace}
\newcommand{\we}{I\xspace}
\newcommand{\We}{I\xspace}
\submitjournal{PASP}

\shorttitle{SNe~Ia with \RST $K$}
\shortauthors{Rubin}

\graphicspath{{./}{figures/}}

\begin{document}

\title{Evaluating $K$ bands for \RomanSpelled Rest-Frame NIR SN~Ia Distances}

\correspondingauthor{David Rubin}
\email{drubin@hawaii.edu}

\author[0000-0001-5402-4647]{D. Rubin}
\affiliation{Department of Physics and Astronomy, University of Hawai`i at M{\=a}noa, Honolulu, Hawai`i 96822, USA}
\affiliation{E.O. Lawrence Berkeley National Laboratory, 1 Cyclotron Rd., Berkeley, CA 94720, USA}

\begin{abstract}

Recently, the \RomanSpelled (\RST) Project raised the possibility of adding another filter to \RST. Based on the Filter Working Group's recommendations, this filter may be a $K$-band filter, extending significantly redder than the current-reddest $F184$. Among other scientific possibilities, this $K$ filter raises the possibility of measuring SNe~Ia in the rest-frame NIR out to higher redshifts than is possible with the current filter complement. \We perform a simple survey optimization for NIR SN~Ia distances with \RST, simultaneously optimizing both filter cutoffs and survey strategy. \We find that the roughly optimal $K$ band extends from 19,000\ang--23,000\ang (giving exposure times roughly half that of a 20,000\ang--23,000\ang $K_s$ filter). Moving the $K$ much redder than this range dramatically increases the thermal background, while moving the $K$ band much bluer limits the redshift reach. Thus \we find any large modification reduces or eliminates the gain over the current $F184$. \We consider both rest-frame $Y$ band and rest-frame $J$ band surveys. Although the proposed $K$ band is too expensive for a large rest-frame $Y$ band survey, it increases the rest-frame $J$ Figure of Merit by $59\%$.

\end{abstract}


\section{Introduction} \label{sec:intro}

The \RomanSpelled (\RST) is a 2.4 meter space telescope designed for coronagraphy and wide-field optical-NIR imaging and slitless spectroscopy scheduled for launch in the mid 2020's \citep{spergel15}. The \RST Wide-Field Instrument (WFI) will observe 0.28 square degrees per pointing with $0\farcs 11$ pixels. The large field of view ($> 200$ times that of the {\it Hubble Space Telescope}, \HST, Wide-Field Camera 3 IR), infrared sensitivity, and spectroscopic capability make \RST a powerful complement to optical cosmological surveys such as that carried out by the Vera C. Rubin Observatory \citep{VRO}.

\begin{figure}
    \centering
    \includegraphics[width=0.6\textwidth]{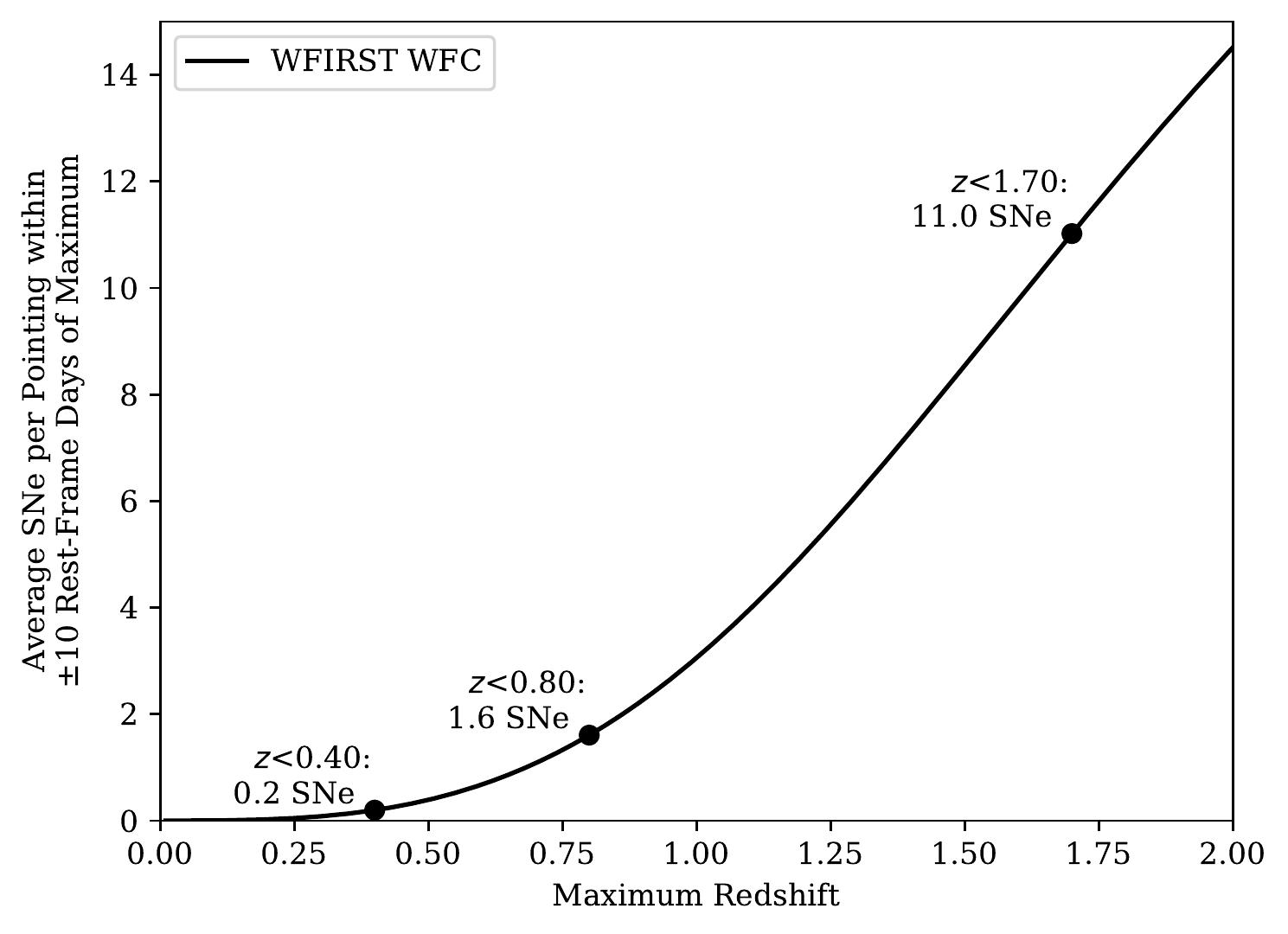}
    \caption{Average number of SNe~Ia within 10 rest-frame days of maximum (20~rest-frame day window) in a random WFI pointing as a function of maximum redshift (the distribution is cumulative in redshift). \We assume a SN~Ia volumetric rate from \citet{rodney14}. Multiplex is a factor above $z \sim 0.6$ (this value depends on the exact phase range considered useful). Below this redshift, targeted followup observations can improve survey efficiency.}
    \label{fig:multiplex}
\end{figure}

One of \RST's cosmological probes is Type Ia supernovae (SNe~Ia), to be observed in large ($\sim$~10's of square degrees), cadenced, deep fields \citep{spergel15, hounsell18}. Figure~\ref{fig:multiplex} shows that \RST WFI will be large enough to capture many SNe~Ia at once in a rolling survey above $z \sim 0.6$.

\begin{figure}
    \centering
    \includegraphics[width=0.8\textwidth]{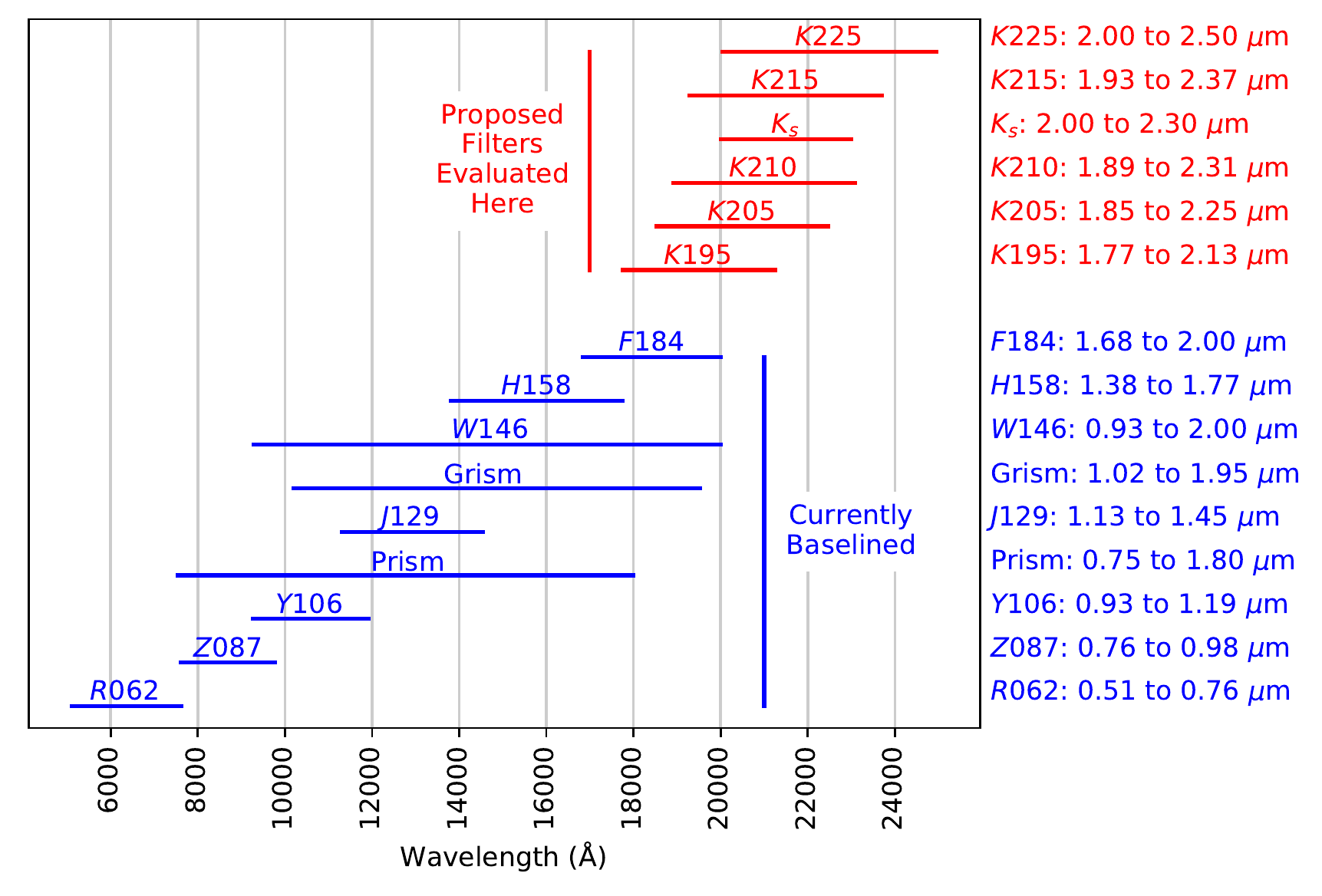}
    \caption{Spectral range of each WFI spectral element (50\% on to 50\% off). The bottom group shows elements that are currently baselined, while the top group shows possible $K$ filters evaluated in this work.\label{fig:spectral} }
\end{figure}

Figure~\ref{fig:spectral} shows the WFI spectral elements (filters, grism and prism), as well as the possible redder filters considered in this work (referred to as ``$K$''). Each filter is labeled with its central wavelength in a three-digit code similar to \HST (e.g., the $F184$ filter has a central wavelength of 1.84 $\mu$m). The baselined filter set was chosen to span most of the sensitive wavelength range of the detectors (H4RG's with a 2.5~$\mu$m cutoff); the 2.0~$\mu$m cutoff of the reddest baselined filter ($F184$) minimizes thermal background, thus taking advantage of the dark sky possible in space.

Recently, the \RST Project raised the possibility of adding another filter to \RST. Based on the Filter Working Group's recommendations, this filter may be a $K$-band filter, extending significantly redder than the current-reddest $F184$. As SNe~Ia are intrinsically more standard in the rest-frame NIR (e.g., \citealt{meikle00, krisciunas05, bn12, avelino19, mandel20}) and possibly have lower dust extinction systematic uncertainties, it is useful to investigate using and optimizing this new $K$-band filter for measuring SNe~Ia. 

\section{Survey Simulations and Optimization} \label{sec:simulations}

As much as possible, \we make survey simulation assumptions that are likely to be correct in a relative sense, and thus any uncertainties cancel out of the relative FoM/survey optimization. For example, when targeting a certain redshift range, the exposure time is set such that the median SN~Ia (on the median host-galaxy background) at the far end of that range has S/N 10 at maximum per 2.5 rest-frame days. This corresponds to a S/N of 10 with a cadence of 5 observer-frame days when targeting $z= 1$, but when targeting $z=0.4$ (3.6 rest-frame days cadence) a S/N of 12 ($10 \sqrt{3.6/2.5}$). If a detailed analysis showed that S/N 8 was a better assumption, the exposure times could be scaled to $\sim (8/10)^2$ while all surveys would contain $\sim (10/8)^2$ or 56\% more SNe, but all the surveys would scale the same way relative to each other (except for the nearby SNe~Ia and overheads, which are described below). \We explicitly verify that the conclusions do not change with total survey time, as discussed below.

\begin{figure}
    \centering
    \includegraphics[width = 0.6\textwidth]{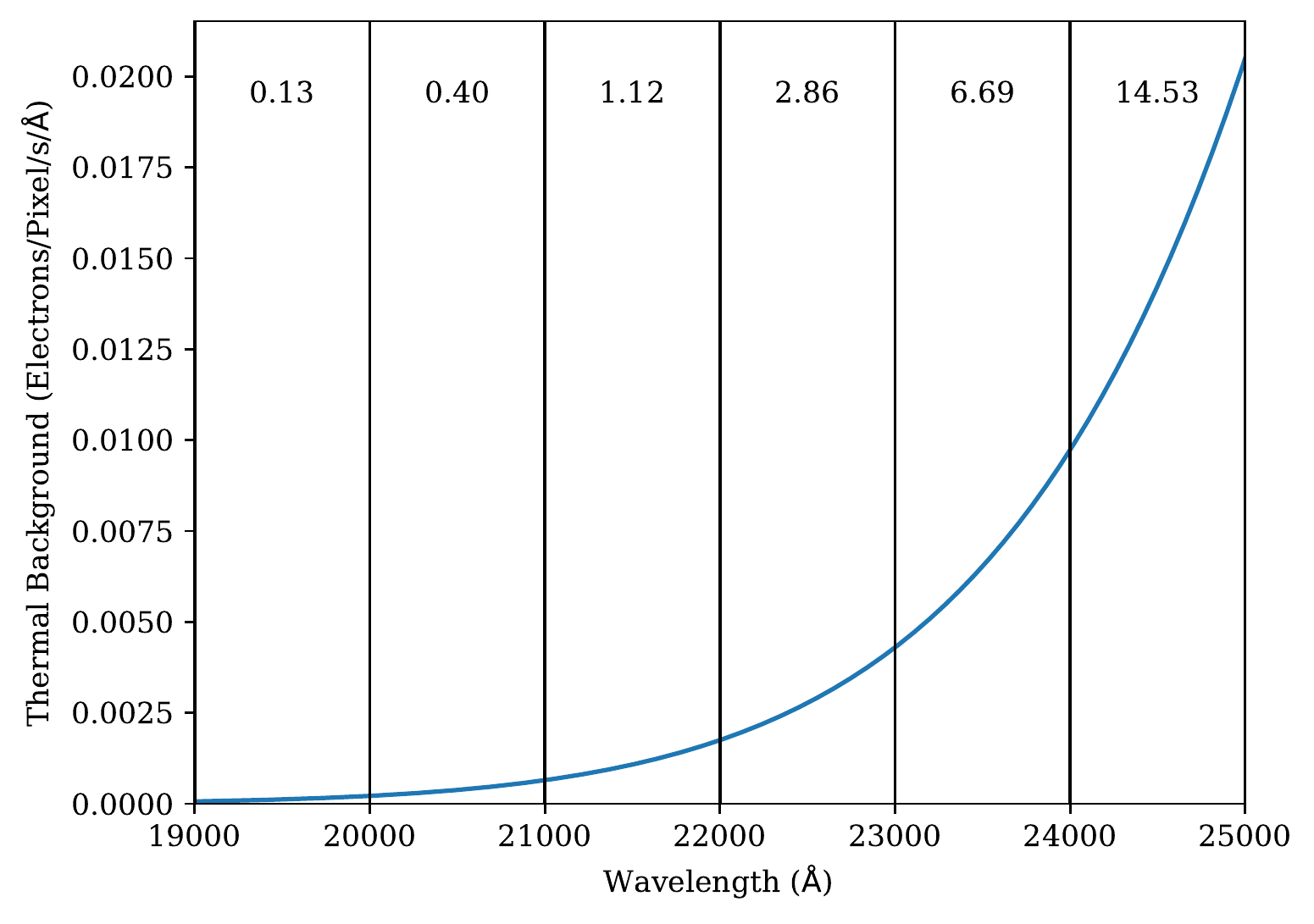}
    \caption{Thermal background (in e$^-$/pixel/s/\ang) as described in the text. Also shown is the total in bins of 1,000\ang.}
    \label{fig:thermal}
\end{figure}

By using this same argument that all surveys being optimized scale together, \we do not include any source of systematic uncertainties in the forecast. \We do note that controlling many sources of systematic uncertainty should be possible, including:
\begin{itemize}
    \item Photometric nonlinearity in the $K$ band, possibly calibratable using lamp-on/lamp-off with the spectral tail of the reddest \RST Relative Calibration System LEDs, e.g., \citet{dejong06}
    \item Uncertainties in the NIR SED model for SNe Ia, possibly addressable with a combination of \RST prism spectrophotometry, and other programs such as ground-based spectra \citep[e.g.,][]{hsiao19} and space-based spectra (e.g., Supernovae in the Infrared avec Hubble, \HST~GO~15889 PI:Jha)
    \item Flatfielding uncertainties, especially as most of the thermal emission will be near the filter cutoff wavelength and not near the central wavelength of the filter (see Figure~\ref{fig:thermal}). Much of this uncertainty will average out as SNe are distributed randomly over the focal plane.
    \item Photometric-classification uncertainties (these uncertainties are expected to be small with double-peaked rest-frame NIR SN~Ia light curves).
    \item $K$-band standard-star SED uncertainties, which may be addressable with {\it James Webb Space Telescope} spectrophotometry.
    \item Selection effects, which should be small given the high S/N of the light curves (these surveys effectively target a volume-limited sample defined by a given rest-frame NIR range)
\end{itemize}
Finally, \we do not consider alternatives and synergies with NIR light curves for controlling astrophysical systematic uncertainties (e.g., prism spectrophotometry with a SN~Ia subclassification approach such as \citealt{fakhouri15}). A detailed consideration of this is simply beyond the scope of this work.

The survey optimization assumes a Hubble-Lema\^{i}tre diagram dispersion of 0.12 magnitudes (c.f., \citealt{avelino19, mandel20}), added in quadrature with $0.055 z$ lensing dispersion \citep{jonsson10}, with both dispersions assumed Gaussian. \We assume 800 nearby SNe ($0.12/\sqrt{800} = 0.004$ mag) calibrated on the same photometric system as \RST. \We assume that, given the modest redshifts of these SNe ($z \lesssim 1$), ground-based facilities (e.g., \citealt{pfs, desi, fourmost}) are able to obtain the redshifts of the host galaxies or live SNe. \We compute the Dark Energy Task Force Figure of Merit (DETF FoM, \citealt{detf}) using a 0.2\% CMB shift-parameter constraint assuming a flat universe. This is the same FoM that was used for the SN survey in \citet{spergel15}.

My survey simulations use a simple pixel-level optimal extraction using WebbPSF \RST PSFs \citep{perrin14}. \We use 2.55 square meters for the peak effective area of the proposed $K$ filters, and the \RST Project-supplied effective areas for the other filters.\footnote{\url{https://roman.gsfc.nasa.gov/science/201907/WFIRST_WIMWSM_throughput_data_190531.xlsm}} \We use the zodiacal background of \citet{aldering02}, appropriate for a high ecliptic latitude of $\pm 75^{\circ}$. The supernova fluxes use SALT2-Extended\footnote{2013 SNANA version called through SNCosmo}; the galaxy backgrounds use a model trained on the real backgrounds of high-redshift SNe \citet{riess07, riess18}, described in more detail in \citet{rubin21}. \We assume a conservative 20~e$^-$ of read noise per 2.825-second read and a 5~$e^-$ floor and a subdominant 0.015 e$^-$/pixel/s of dark + stray-light background.\footnote{\We use a simple quadrature sum of $\sqrt{N}$ Poisson noise and read noise for the noise per pixel, which is not quite accurate \citep{vacca04, rauscher07}. \We am thus implicitly assuming a more optimal weighting of the detector readouts (e.g., \citealt{kubik16}). However, \we am assuming conservative values for the read noise, so this quadrature sum should be achievable with a more detailed treatment.} The thermal model (which is tuned to produce very similar thermal backgrounds as given by the \RST Project) assumes 264K blackbody emission from a combined 2.4~meter primary and secondary mirror with total throughput-weighted emissivity of 0.03. Combined with 10~$\mu$m $0\farcs11$ square pixels (18.75~meter focal length), this gives the model shown in Figure~\ref{fig:thermal}.

Table~\ref{tab:expsoure} shows how to translate a given redshift distribution of SNe into exposure times. A trial SN survey is specified in terms of the relative number of SNe in each redshift bin. This number of SNe is translated into the areas and depths necessary to measure that many SNe, with any SNe at lower redshift coming for free. \We consider surveys measuring at least as red as rest-frame $Y$ band ($\lambda \gtrsim 1.04 \mu$m) and surveys measuring at least as red as rest-frame $J$ band ($\lambda \gtrsim 1.25 \mu$m). \We do not consider rest-frame $H$-band surveys ($\lambda \gtrsim 1.635 \mu$m); even with the reddest practical filters; these would be limited to $z \lesssim 0.4$. Figure~\ref{fig:multiplex} shows that these surveys would be more practical with a targeted followup approach (e.g., RAISIN \HST~GO~13046 PI: Kirshner), which is beyond the scope of the rolling surveys considered in this work.

\We consider five survey variants for both rest-frame $Y$ and rest-frame $J$ surveys:
\begin{itemize}
    \item There is no additional $K$-band filter, so the survey must use the existing $H158$ and $F184$ filters.
    \item The optimum $K$ filter, which is optimized simultaneously with the survey. For example, a rest-frame $Y$ survey targeting a maximum redshift of 1 uses the $K205$ filter (see Table~\ref{tab:expsoure}). The rest-frame wavelength and maximum redshift set the observer-frame central wavelength (with rounding); the filter width around this wavelength is set to maximize S/N (specifically, \we maximize filter~width/$\sqrt{\mathrm{thermal\ background}}$). The $K$ filters from Table~\ref{tab:expsoure} are shown in the top group of Figure~\ref{fig:spectral}.
    \item \We optimize the survey, but the only $K$ filter considered is the $K210$. As \we show, this filter is roughly optimal for rest-frame $J$-band measurements.
    \item As above, but using $K215$. This filter is forced to a similar effective wavelength to the $K_s$ filter, but has a width chosen to maximize S/N.
    \item As above, but using $K_s$ (the filter proposed by \citealt{sfauffer18}).
\end{itemize}

After computing the relative numbers of pointings required to achieve the SN numbers, \we scale the relative numbers to absolute numbers by requiring that the total imaging take 0.2~years or 40\% of the total SN survey time (0.5~years). As described above, my conclusions are not sensitive to the assumption of 0.2 years, as all the surveys scale in the same way with total survey exposure time (\we verify this by optimizing surveys with other total times). \We also verify that my results are similar with a 260K (rather than 264K) telescope. \We optimize each survey with a downhill-simplex code \citep{neldermead} subject to the constraint that the SNe in each redshift bin must be nonnegative. Downhill-simplex minimization can be challenging even in moderate numbers of dimensions, so \we assure convergence by restarting the minimization many times (both in parallel, with randomly chosen starting conditions, and serially, with another maximization started where the previous one terminates).

\begin{deluxetable*}{c|cc|cc|cc|cc||cc|cc|cc|cc}
\rotate
\tablewidth{0pt}
\tablecaption{Exposure Times and Filters for Median SN as a Function of Maximum Redshift\label{tab:expsoure}}
\tablehead{
\colhead{Maximum} & \multicolumn{8}{c}{Rest-Frame $Y$ Surveys}  & \multicolumn{8}{c}{Rest-Frame $J$ Surveys} \\
\colhead{Redshift} & \multicolumn{2}{c}{Optimum Filter}  & \multicolumn{2}{c}{$K210$} & \multicolumn{2}{c}{$K215$}  & \multicolumn{2}{c}{$K_s$} & \multicolumn{2}{c}{Optimum Filter}  & \multicolumn{2}{c}{$K210$} & \multicolumn{2}{c}{$K215$}  & \multicolumn{2}{c}{$K_s$}
}
\startdata
0.2 & $H158$ & 22.6 & $H158$ & 22.6 & $H158$ & 22.6 & $H158$ & 22.6 & $H158$ & 22.6 & $H158$ & 22.6 & $H158$ & 22.6 & $H158$ & 22.6 \\ 
0.3 & $H158$ & 39.6 & $H158$ & 39.6 & $H158$ & 39.6 & $H158$ & 39.6 & $F184$ & 79.1 & $F184$ & 79.1 & $F184$ & 79.1 & $F184$ & 79.1 \\ 
0.4 & $H158$ & 59.3 & $H158$ & 59.3 & $H158$ & 59.3 & $H158$ & 59.3 & $F184$ & 104.5 & $F184$ & 104.5 & $F184$ & 104.5 & $F184$ & 104.5 \\ 
0.5 & $H158$ & 79.1 & $H158$ & 79.1 & $H158$ & 79.1 & $H158$ & 79.1 & $F184$ & 141.2 & $F184$ & 141.2 & $F184$ & 141.2 & $F184$ & 141.2 \\ 
0.6 & $F184$ & 194.9 & $F184$ & 194.9 & $F184$ & 194.9 & $F184$ & 194.9 & $K205$ & 494.4 & $K210$ & 810.8 & $K215$ & 1387.1 & $K_s$ & 1604.6 \\ 
0.7 & $F184$ & 274.0 & $F184$ & 274.0 & $F184$ & 274.0 & $F184$ & 274.0 & $K210$ & 1118.7 & $K210$ & 1118.7 & $K215$ & 1904.1 & $K_s$ & 2132.9 \\ 
0.8 & $F184$ & 350.3 & $F184$ & 350.3 & $F184$ & 350.3 & $F184$ & 350.3 & $K225$ & 7200.9 & \nodata & \nodata & \nodata & \nodata & \nodata & \nodata \\ 
0.9 & $K195$ & 709.1 & $K210$ & 2500.1 & $K215$ & 3960.7 & $K_s$ & 4429.6 & \nodata & \nodata & \nodata & \nodata & \nodata & \nodata & \nodata & \nodata \\ 
1.0 & $K205$ & 1952.1 & $K210$ & 3220.5 & $K215$ & 5231.9 & $K_s$ & 6483.4 & \nodata & \nodata & \nodata & \nodata & \nodata & \nodata & \nodata & \nodata \\ 
1.1 & $K215$ & 6378.9 & \nodata & \nodata & $K215$ & 6378.9 & $K_s$ & 8232.1 & \nodata & \nodata & \nodata & \nodata & \nodata & \nodata & \nodata & \nodata \\ 
1.2 & $K225$ & 17664.7 & \nodata & \nodata & \nodata & \nodata & \nodata & \nodata & \nodata & \nodata & \nodata & \nodata & \nodata & \nodata & \nodata & \nodata \\ 
\enddata
\tablecomments{Assumptions for computing required exposure time as a function of maximum redshift in a field. In addition to the exposure times listed above, \we assume a 70.625s slew time per pointing (this time is quantized in term of the 2.825s readout time of the WFI detectors, which read and reset during slews for stability and to avoid persistence slewing across bright stars). The left columns show filters and exposure times for surveys that target rest-frame $Y$ band; the right columns show surveys that target rest-frame $J$ band.}
\end{deluxetable*}

\begin{deluxetable*}{c|cc|cc}
\rotate
\tablewidth{0pt}
\tablecaption{Optimized surveys, filter choices, and FoM values\label{tab:results}}
\tablehead{
\colhead{Variant} & \multicolumn{2}{c}{Rest-Frame $Y$ Surveys}   & \multicolumn{2}{c}{Rest-Frame $J$ Surveys}  }
\startdata
No $K$	&	25 deg$^2$ $H158$ 79.1 s, 20 deg$^2$ $F184$ 350.3 s	&	267	&	57 deg$^2$ $F184$ 141.2s	&	68	\\
Optimum $K$	&	31 deg$^2$ $H158$ 79.1 s, 9.5 deg$^2$ $K195$ 709.1 s	&	285	&	24 deg$^2$ $F184$ 141.2 s, 6.0 deg$^2$ $K210$ 1118.7 s	&	108	\\
$K210$	&	25 deg$^2$ $H158$ 79.1 s, 20 deg$^2$ $F184$ 350.3 s	&	267	&	24 deg$^2$ $F184$ 141.2 s, 6.0 deg$^2$ $K210$ 1118.7 s	&	108	\\
$K215$	&	25 deg$^2$ $H158$ 79.1 s, 20 deg$^2$ $F184$ 350.3 s	&	267	&	28 deg$^2$ $F184$ 141.2 s, 3.1 deg$^2$ $K215$ 1904.0 s	&	86	\\
$K_s$	&	25 deg$^2$ $H158$ 79.1 s, 20 deg$^2$ $F184$ 350.3 s	&	267	&	30 deg$^2$ $F184$ 141.2 s, 2.6 deg$^2$ $K_s$ 2132.9 s	&	82	\\
\enddata
\tablecomments{Results from my optimizations. The first column presents the survey variants \we consider: no $K$, a $K$ filter that is optimized with the survey strategy, the $K210$ filter, the $K215$ filter, and the $K_s$ filter. The next two columns present the optimized survey for each variant and the FoM value for rest-frame $Y$ band distances. Finally, the last two columns present the optimized survey for each variant and the FoM value for rest-frame $J$ band distances.}
\end{deluxetable*}

\section{Results and Conclusions} \label{sec:conclusions}

Table~\ref{tab:results} presents the optimized surveys, filters, and NIR-only FoM values. For the rest-frame $Y$ surveys, using $K195$ rather than $F184$ can raise the FoM value 7\%. However, $K195$ is not meaningfully redder than the $F184$, and so including $K195$ is hard to justify for a 7\% increase. None of the other $K$ filters considered supplant the $F184$ (although $K$-band data may be of use for distance cross-checks for a fraction of the SNe observed in the bluer filters, e.g., \citealt{dhawan18, burns18, mandel20}). For the rest-frame $J$ surveys, including $K210$ is highly preferred, raising the FoM by 59\% compared to an optimized survey without $K$. The $K215$ filter only offers a 26\% increase in FoM, and the $K_s$ filter only offers a 21\% increase in FoM.

For measuring SNe~Ia in the NIR, $K210$ thus offers the optimum combination of minimizing thermal background and measuring as red as practical. It is not surprising that the $K210$ filter is significantly more optimal that $K_s$; it gathers $\sim 4/3$ the light (by spanning 19,000\ang--23,000\ang to the 20,000\ang--23,000\ang range of $K_s$) with the same level of thermal background (same 23,000\ang cutoff), and so needs only $\sim (3/4)^2 = 56\%$ the exposure time.

\acknowledgments

This work was supported by NASA through grant NNG16PJ311I (Perlmutter \RST Science Investigation Team). \We acknowledge careful feedback from Greg Aldering, Ben Rose, Susana Deustua, and Rebekah Hounsell.

\software{
Astropy \citep{astropy},
Mathematica \citep{Mathematica},
Matplotlib \citep{matplotlib}, 
Numpy \citep{numpy}, 
Python \citep{python3}, 
SciPy \citep{scipy},
SNCosmo \citep{sncosmo}
}

{}

\end{document}